\documentclass[prb,aps,onecolumn,groupedaddress,floats,showpacs,final,12pt]{revtex4}
\usepackage{amsmath}
\usepackage{graphicx}
\usepackage{dcolumn}
\usepackage{bm}
\usepackage{color}
\usepackage{ulem}
\definecolor{blue}{rgb}{0.3,0.3,0.9}

\def\beq{\begin{eqnarray}}
\def\eeq{\end{eqnarray}}

\begin{document}

\author{Victor Fleurov$^1$ and Anatoly B. Kuklov$^2$}
\affiliation{$^1$ Raymond and Beverly Sackler Faculty of Exact Sciences, School of Physics and Astronomy, Tel-Aviv University -
Tel-Aviv 69978, Israel }
\affiliation{$^2$ Department of Physics and Astronomy, CSI, and the Graduate Center of CUNY, New York.}

\title{Cooperative phase transitions in the system of photons and dye molecules}

\date{\today}
\begin{abstract}
Bose condensed light can form new phases [\onlinecite{njpus}] in a dye filled cavity due to the presence of the orientational disorder created by dye molecules  which are essentially frozen on the time scale of the photonic thermalization (few ps). At longer times (few ns) molecular degrees of freedom -- orientations and positions -- become important. Including them on equal footing with photons can change the nature of the photonic condensation -- it can proceed as Ist order phase transition which can also result in the mutual phase separation effect -- for photons and dye. The analysis is conducted within the mean field approach in the thermodynamic limit. Recommendations for the experimental detection of the transition nature are formulated.
\end{abstract}


\maketitle

\section{Introduction}
Bose Einstein condensate (BEC) as a fully equilibrium phase can only occur in an ensemble of conserved bosons. Thus, photons cannot form BEC. 
However, if the processes of photon creation and disappearance are characterized by time scale much longer than the time of the photon thermalization, the photonic BEC can form as an equilibrium phase for all practical purposes. This logic has guided the realization of the photonic BEC in the resonant cavity 
filled with dye molecules   [\onlinecite{Nature}] (see also in Refs.[\onlinecite{,Schmitt,NS14,MN15,GPO18,Weitz2017}]). 

In this system the main role of the molecules is to thermalize light on a time scale (of few ps) which is much shorter than the time of the photon losses (several ns). Another important aspect of the experiment [\onlinecite{Nature}] is a significant difference between typical photon energies (about $\hbar\omega_0 \approx $2 eV) and temperature $T$ of the cavity (about 0.03 eV). This essentially eliminates thermal processes leading to photon non conservation (controlled by the factor $ \exp(-\hbar\omega_0/T) \sim 10^{-30}$). Accordingly, photons can be treated as conserved particles characterized by finite chemical potential $\mu_0$ determined by external pumping [\onlinecite{Nature,Schmitt,NS14,MN15,GPO18}]. In addition, the quasi 2D geometry of the cavity  [\onlinecite{Nature}] introduces dimensional quantization of the photonic modes which gives photons an effective mass $m \approx \hbar\omega_0/c^2$, where $c$ is speed of light. Thus, a cavity photon becomes a massive particle which is a factor $10^5-10^6$ lighter than electron. This allows achieving the degeneracy condition (when the distance between particles becomes comparable to their thermal wavelength) at room temperature for 2D densities of photons as low as $\sim 10^{11}$m$^{-2}$ (if compared with typical 2D densities $\sim 10^{19}$m$^{-2}$ of solids). This estimate, however, does not take into account the presence of the dye molecules. As it will be discussed below (see in Sec.\ref{sec:normal}), under certain condition these molecules can carry most of the normal excitations, and this changes the condensation condition -- from what is traditionally known for ideal gas of bosons.   

It is important to realize that the role of the dye molecules is not limited by the thermalization of photons. As discussed in Refs.[\onlinecite{Sela-14,SFY16}],
the dye molecules (each treated as a two-level system (TLS)) can interact with each other by exchanging virtual photons similarly to the F\"orster mechanism of the energy transfer (see in Ref.[\onlinecite{Forster}]). In the quasi-2D geometry this may result in the formation of a collective state of the molecules even with no photonic condensate present. 

As noted in Refs.[\onlinecite{njpus}], virtual electronic transitions within each molecule introduce anisotropy lowering the symmetry of the photonic order parameter from O(4) to O(2)$\times$Z$_2$ (see also in Ref.[\onlinecite{Rubo-07}]). This guarantees formation of the algebraic photonic condensate and the photonic superfluid (PSF). Alternatively, without such a lowering of the symmetry, no algebraic order of the photonic condensate would be possible according to the Conformal Field Theory in 2D [\onlinecite{Shenker}]. 

Another crucial aspect is that molecular degrees of freedom -- their orientations and positions -- are essentially classical variables at temperatures relevant for the experiment.  One consequence of this is the possibility of the formation of so called {\it geometrically} paired photonic condensate [\onlinecite{njpus}].       
This phase is possible as a transient phenomenon on the time scale characterized by three orders of magnitude -- from few ps to few ns  when the dye molecules can be treated as a source of static disorder. [Such a phase can be made an equilibrium one by imposing additional optical fields  [\onlinecite{njpus}]). At longer time scales the rotational and translational molecular degrees of freedom come into play. Thus, the analysis of the photonic condensation should include these variables too -- in addition to the optical fields. Here we focus on time scales much longer than the photonic thermalization.

As discussed below, virtual electronic transitions between ground and excited states of each TLS stimulated by the presence of the condensing photons induce preferential orientation of the molecules along the optical field. As a result, under certain conditions the transition can become discontinuous into a linearly polarized photonic condensate -- in contrast to the short time situations analyzed in Ref.[\onlinecite{njpus}]. Initially, the system is considered in the limit when spatial distribution of the dye is uniform and fixed as an external parameter, while the orientation of the molecules is allowed to relax in the local field of the photonic condensate. In this approximation the ensemble is grand canonical with respect to the number of photons. Then, in order to consider full equilibrium, the grand canonical description is extended to the number of dye molecules, which allows their density fluctuations to occur. Accordingly, the discontinuous nature of the transition leads to the phase separation effect. This important aspect of the system has already been emphasized in Ref.[\onlinecite{arxiv}], although without including the orientational relaxation.

\section{Dynamical degrees of freedom and the Hamiltonian}
\begin{figure}[!htb]
	\includegraphics[width=1.0 \columnwidth]{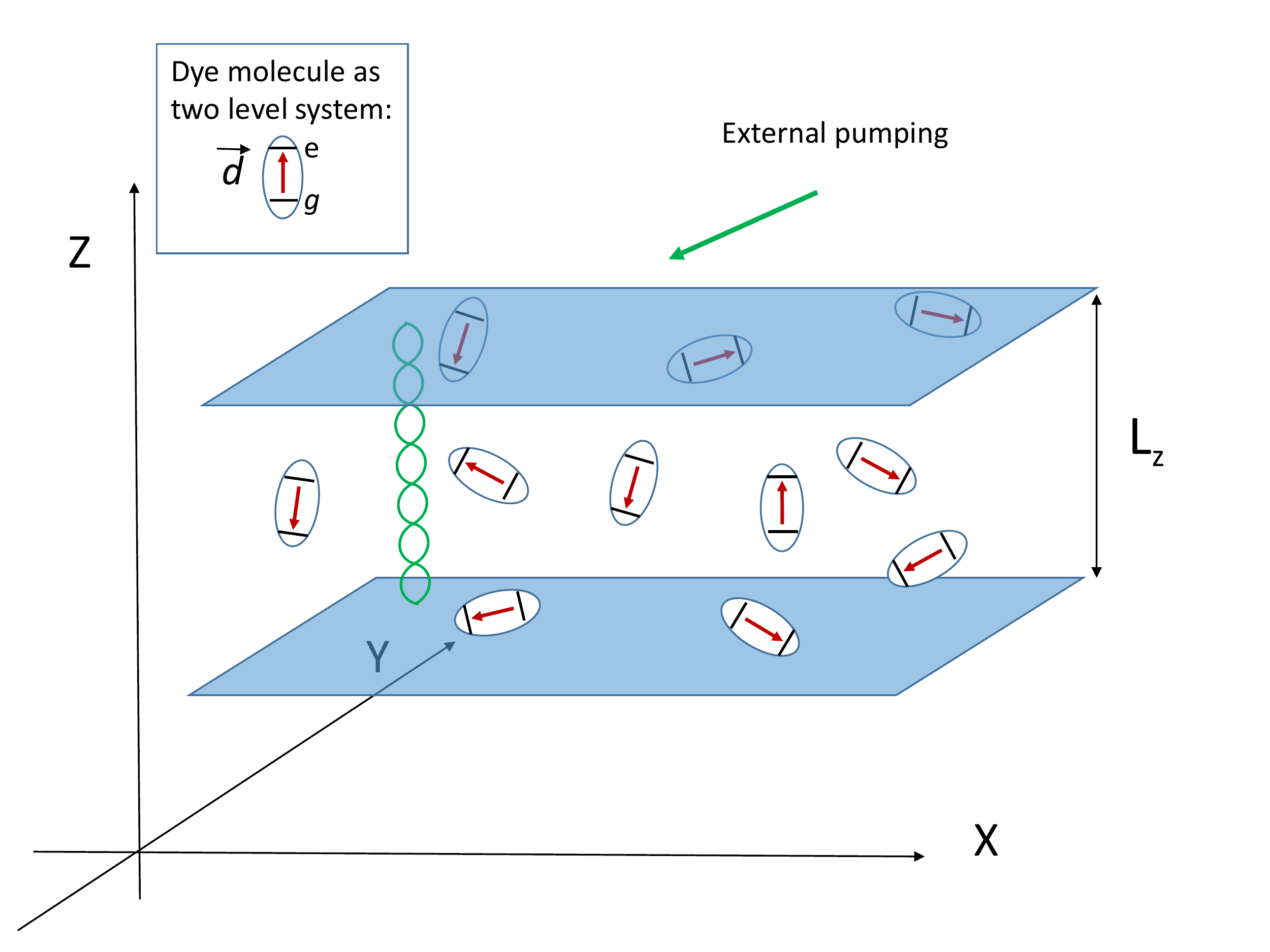}
	\vskip-8mm
	\caption{(Color online) Sketch of the setup [\onlinecite{Nature}]. Two parallel planes [\onlinecite{note}] depict cavity mirrors with the dye solution (randomly oriented ellipses with arrows) in between. 
 Externally pumped light forms standing wave along Z-direction (green chain of ellipses). Low energy perturbations of the photonic condensate are characterized by small momenta $\vec{k}=(k_x,k_y,0)$ with the X,Y components only. Inset:  dye molecule modeled as two level system characterized by the ground ({\it g}) and excited (e) states and the dipole moment $\vec{d}$ (shown by arrow) of the $S_0 \to S_1$ transition.}
	\label{fig_set}
\end{figure}

 The setup [\onlinecite{Nature}] is schematically explained in Fig.~\ref{fig_set}. There are several relevant degrees of freedom: 1. Electric field of the Bose condensed photons; 2. Electronic states of dye molecules; 3. Orientation of the molecules determining the direction of the dipolar transition $S_0 \to S_1$ responsible for the absorption and emission of photons; 4. Center of mass positions of the dyes; 5. Vibronic deformations of the dye molecules responsible for the thermalization of the photons. 

It is important to keep in mind the hierarchy of time scales. The shortest one is the thermalization time $\tau_t \sim 20$ps [\onlinecite{Nature}]. It is about 2-3 orders of magnitude shorter than the life time of excitations $\tau_e$ (of several ns) in the cavity [\onlinecite{Shafer,Nature}]. Thus, for all practical purposes the system can be treated as an equilibrium one on the time scale longer than $\tau_{t}$ in the limit of weak pumping. The translational and rotational dynamics of the dye subsystem is characterized by its own time scales. These are determined by dynamical viscosity $\eta$ of the solvent  (see in Ref.[\onlinecite{Landau_hydro}]). The rotational diffusion time $\tau_r \sim \eta a^3/T$, where $a\sim$1nm is a typical size of the dye molecule  (Boltzmann constant is taken as unity). For typical viscosity $\eta \sim 10^{-3}$Pa$\cdot$s, this time is $\tau_r \sim$ 1-10 ns.  Thus, the orientation of a dye molecule defining direction of its dipolar moment of the electronic transition $\vec{d}$ is, practically,  frozen and random on the time scale $t << \tau_r$. Beyond this time the dipolar variable $\vec{d}$ must be treated as a dynamical (diffusive) degree of freedom.  The translational diffusion of the dye molecules is characterized by a spread of times -- from the travel time $\tau_{trc}\sim \eta R_n^2 a/T$ to its closest neighbor (about $R_n \sim $10nm apart) to the time $\tau_{cl}\sim \eta R_{cl}^2 a /T$ it takes to diffuse on the spatial scale $R_{cl}$ set by the cavity geometry. Taking $R_{cl} \sim 1-10\mu$m (comparable to the size of the condensate in Ref.[\onlinecite{Nature}]), this range covers from $\tau_{trc} \sim 10^{-7}-10^{-6}$s to $\tau_{cl}\sim 0.1-1$s.   Similarly to the rotational degree of freedom, center of mass positions of the molecules should be considered as dynamical (diffusive) degrees of freedom on times longer than $\tau_{trc}$. In Ref.[\onlinecite{njpus}], the analysis has been performed at time scales much shorter than $\tau_r$ . 

Our focus is on evaluating equilibrium free energy by integrating out all relevant degrees of freedom but photons -- in order to obtain the effective Gross-Pitaevskii action for the condensate and its low energy excitations. 
The order parameter of light is a complex vector $\vec{E}=\vec{E}(\vec{x},z)$ representing the amplitude of electric field in the rotating wave approximation, with $\vec{x}=(x,y)$ and $z$ being coordinates along and perpendicular to the XY plane (see Fig.\ref{fig_set}) of the resonator [\onlinecite{Nature}], respectively. 
The geometry of the cavity in the experiment [\onlinecite{Nature}] resembles a rectangular waveguide [\onlinecite{note}] made of two parallel mirrors separated by a distance $L_z \sim 2 \mu$m (see Fig~\ref{fig_set}). The cavity supports two types of modes (see in e.g. [\onlinecite{J98}]). The TM mode has a nonzero electric field component  along the resonator Z-axis, whereas it is zero for the TE mode.  Correspondingly the Z component of the magnetic field is zero for the TM mode and nonzero for the TE mode. The components of $\vec{E}$ along the XY plane in both cases can be represented as
\beq
\vec{E}=\phi(z) \vec{\psi}(\vec{x}),\,\, \phi(z)= \phi_0\sin(q_0 z),\,\, \phi_0=\sqrt{\frac{2\hbar \omega_0}{\varepsilon_0L_z}}
\label{anz}
\eeq
where $\phi(z)$ describes the standing wave along the Z-axis, with $q_0$ determined by the vertical dimension $L_z$ of the cavity as $q_0=7\pi/L_z$ [\onlinecite{Nature}] (see in Fig.\ref{fig_set}); the complex vector $\vec{\psi}=(\psi_x, \psi_y)$ accounts for the transverse order and its long wave fluctuations in the XY-plane of the resonator. The Z component of $\vec{E}$ is chosen as to ensure the divergenceless nature of the electric and magnetic fields. 
It, however, can be ignored for the purpose of deriving the effective action in the long wave limit and analyzing the nature of the emerging phases and phase transitions of light and dye molecules within the mean field approach. 
In Eq.(\ref{anz}) the normalization in which $\vec{\psi}^\dagger \vec{\psi}$ is the operator of the 2D density of photons is used [\onlinecite{com}], with $\varepsilon_0$ denoting the background dielectric permittivity of the medium.

Each dye molecule described as TLS is characterized by its center of mass position $(x_i,y_i,z_i)$, where the index $i$ runs over all $N$ such molecules, and by the orientation given by the unit vector $\vec{n}_i$ along $\vec{d}$. As emphasized in Ref.[\onlinecite{njpus}], it is important that this transition ($S_0 \to S_1$ characterized by real matrix element $d$) is non-degenerate which leads to lowering the symmetry from O(4) to O(2)$\times$Z$_2$, and, thus, allows the condensation to occur in 2D geometry in the algebraic sense. At time scales longer than few ns the vector field $\vec{n}_i$ becomes classical order parameter.

In the Hamiltonian (see in Refs. [\onlinecite{Keeling_2013,Keeling_2015}])
\beq
H=H_{\rm ph} + H_{\rm tls} + H_v + H_{\rm int} 
\label{Hfull}
\eeq
 the first term 
\beq
H_{\rm ph}=\int d^2x  \frac{\hbar^2}{2m} \nabla_i \psi^\dagger_j \nabla_i \psi_j 
\label{U0}
\eeq
accounts for free photons inside the cavity with the reference zero point set for photons with zero component of the momentum along the XY plane. The term 
\beq
H_{tls} =\sum_{i}\left[ \delta \cdot (\sigma_z)_i + D \vec{E}_i^\dagger \vec{E}_i (\sigma_0)_i\right]
\label{zigma}
\eeq
gives the diagonal part of the TLS Hamiltonian,  with
$\delta = (\epsilon_0 - \hbar \omega_0)/2 >0$ standing for the half of the detuning of the TLS energy $\epsilon_0$ from the energy $\hbar \omega_0$ of the condensed photonic mode, and $(\sigma_z)_i,\, (\sigma^+)_i, \, (\sigma^-)_i$ are the standard Pauli matrices assigned for $i$th TLS.  
The term $\sim D$ is the so called diamagnetic contribution guaranteeing that no spontaneous superradiant transition is possible [\onlinecite{nogo}] (see also Ref.[\onlinecite{nogo2}]). It can be estimated as $D\approx {\rm e}^2/(m_e \omega_0^2)$, where $\rm e$ and $m_e$ stand for electron charge and mass, respectively.
This term is defined at the locations of the TLS centers (and is multiplied by the Pauli unity operator $(\sigma_0)_i$ of the $i$th TLS).  
The part $H_v$ accounts for shape fluctuations of the TLS centers resulting in variations of $\delta $. In a simplified version of the model [\onlinecite{Keeling_2013,Keeling_2015}] this part can be written as a Hamiltonian of (non interacting)  quantum harmonic oscillators 
\beq
H_v=\sum_i \left[ \frac{p_i^2}{2m_v} + \frac{m_v\Omega_v^2 q^2_i}{2}\right]
\label{Hv}
\eeq
in terms of the vibronic variable $q_i$ and the corresponding momentum $p_i$. Here  $\Omega_v$ and $m_v$ stand for the vibronic frequency and some effective mass, respectively. 
Finally, the interactions between photons, TLS and vibrons are accounted for by 
\beq
H_{int}=  \sum_{i} \left[s q_i (\sigma_z)_i -(d \vec{n}_i \vec{E} (\sigma^+)_i +H.c.)\right] .
\label{int}
\eeq
Here $s$ is the the Huang-Rhys interaction parameter determining the relative deformation $q_{eq}= 2s/(m_v\Omega_v^2)$ of a TLS center in its ground and excited states; The term $\sim d$ describes photon  absorption and emission in the rotating wave approximation.

\subsection{Conservation law}
The Hamiltonian (\ref{Hfull}) conserves the total number 
\beq
N_{\rm ex}=\int d^2 x \rho= \int d^2x \vec{\psi}^\dagger(\vec{x})\vec{\psi}(\vec{x}) +  \sum_{i} \frac{1}{2}\left[(\sigma_0)_i + (\sigma_z)_i\right] 
\label{Nexc}
\eeq
of photons and TLS molecules in their excited state. This can be explicitly seen after calculating the time derivative 
$\dot{\rho}=[\rho,H]/(i\hbar)$
 of the excitation density operator
\beq
\rho (\vec{x},t)=\vec{\psi}^\dagger(\vec{x})\vec{\psi}(\vec{x}) + \sum_{i} \frac{1}{2}\left[(\sigma_0)_i + (\sigma_z)_i\right] \delta^{(2)}(\vec{x}-\vec{x}_i),
\label{conserv}
\eeq
 which leads to the 2D continuity equation 
$\frac{\partial \rho}{\partial t} + \vec{\nabla} \vec{J} =0$,
where $\vec{J} = \frac{\hbar}{2mi}\left[ \psi^\dagger_j \vec{\nabla} \psi_j - c.c.\right]$
is  2D density of the photon current; $\delta^{(2)}(\vec{x}-\vec{x}_i)$ is 2D delta-function defined in the space $\vec{x}=(x,y)$.

It is important to emphasize that, in the absence of a direct exchange interaction between TLS molecules, the energy transfer is carried by photons only while storage of the energy is due to both -- photons and TLS excitations. [Translational diffusion of dye molecules is too slow to make any significant contribution to the excitation transport].

At this point it should be mentioned that there are processes of excitation losses which are not accounted for by the Hamiltonian (\ref{Hfull}) -- due to non-radiative transitions (see in Ref.[\onlinecite{Shafer}]) and photons escaping from the cavity. However, in the case when establishing local equilibrium is a much faster process than such losses, considering the conservative model is a justifiable approximation.    

\subsection{Polaron transformation}\label{polaron}
As described in Ref.[\onlinecite{Keeling_2013}], it is convenient to perform the polaron transformation of the Hamiltonian (\ref{Hfull}) as $H \to U^\dagger H U$ with $U=\exp[i \sum_i (q_{eq} p_i/2\hbar) (\sigma_z)_i]$. Then, the vibronic variables are transferred from the term (\ref{zigma}) to the interaction term (\ref{int}) as
\beq
H_{int} \to  - \sum_{i} \left[ d \vec{n}_i \vec{\psi}(\vec{x}_i)\phi(z_i) {\rm e}^{-i q_{eq} p_i/\hbar} (\sigma^+)_i +H.c.\right] 
\label{newint}
\eeq
with the inconsequential shift of the total energy by $m_v \Omega_v^2 q_{eq}^2 N_{TLS}/2$, where $N_{TLS}$ is the total number of the TLS centers, and the
representation (\ref{anz}) has been used.

Vibronic structure of excitations of the dye molecules is responsible for quite wide absorption and spontaneous emission bands (see in Ref.[\onlinecite{Shafer}]). The question is how big is the contribution of vibrons to the equilibrium free energy of the system in the presence of the condensate. 

The structure of the term (\ref{newint}) shows that the vibrons redefine the phase of the field $\vec{\psi}$, which can be viewed as a shift $\vec{\psi} \to \vec{\psi} {\rm e}^{-i q_{eq} p_i/\hbar}$. 
Thus, the effect of vibrons consists of introducing quantum and thermal noise on the condensate phase. This noise, however, is spatially uncorrelated (on a scale larger than $\sim 1$ nm) and, thus, it cannot destroy macroscopic (and mesoscopic) coherence of the condensate. This situation
resembles the significant narrowing of the emission line during lasing if compared with the spontaneous emission. Thus, a fare estimate of the vibronic contribution can be obtained from Eq.(\ref{newint}) by replacing the vibronic factor by the Debye-Waller type factor
$\exp(iq_{eq}p_i/\hbar) \to   \exp(-q_{eq}^2\langle p^2\rangle/2)$, where the equilibrium mean $\langle p^2\rangle$ is evaluated over the vibronic Hamiltonian (\ref{Hv}). Typical values of $q_{eq}$ are $\sim 0.02 \AA$ (see in Ref.[\onlinecite{Shafer}]). Since the effective mass $m_v$ is determined by the reduced mass of C-C or C-N pair changing its length upon absorption (see in Ref.[\onlinecite{Shafer}]), one can estimate the Deby Waller factor as $\sim \exp(- 10^{-2}) - \exp(- 10^{-1})$ in the
limit $\hbar \Omega_v \leq T$ for $T\sim 300$K. Thus for practical purposes the contribution of vibrons to the free energy of the condensate can be safely ignored. 
Similar conclusion about the role of vibrons was reached in Ref. [\onlinecite{Axel}].

\section{Free energy of the photons and dye molecules: no translational diffusion }\label{canonical}
In this section the TLS positions will be treated as a background at some given uniform density $\bar{n}$. In contrast, the orientational variables $\vec{n}_i$ will be integrated out. [As discussed above, this approximation is valid on time scales up to $\sim 10^{-7}-10^{-6}$s].

In the presence of macroscopic occupation of photons forming the condensate field $\vec{\psi}$, the quantum nature of $\vec{\psi},\, \vec{\psi}^\dagger$ can be safely ignored. Such an approach is the basis for describing superfluids by classical fields  $\vec{\psi},\, \vec{\psi}^*$ within the Gross-Pitaevskii equation (see in Refs.[\onlinecite{Landau_9,BK}]). This method is applicable in 2D at finite $T$ as well (see in Ref.[\onlinecite{BK}]) despite the absence of the true off diagonal long range order 
(which is replaced by algebraic off diagonal order and which we are loosely referring to as "condensate"). The goal is to obtain the free energy as a functional of $\vec{\psi}, \vec{\psi}^*$ only -- by calculating the trace of $\exp(-\beta H)$ over the electronic and orientational configurations.

Since the actual conserved quantity is $N_{\rm ex}$ (\ref{Nexc}), within the grand canonical description a chemical potential $\mu_0$ of the excitations should be introduced into Hamiltonian (\ref{Hfull}) as $H \to H-\mu_0 N_{\rm ex}$ in the Gibbs factor $\exp(-\beta H)$, where $\beta=1/T$. This adds the term $ - \mu_0 \int d^2x \vec{\psi}^\dagger(\vec{x})\vec{\psi}(\vec{x})$ to the free photon part, Eq.(\ref{U0}), as well as the term $ -(\mu_0/2) \sum_i[  (\sigma_z)_i+(\sigma_0)_i]$ to the diagonal part of the  TLS Hamiltonian, Eq.(\ref{zigma}).  

As mentioned above, in the presence of the condensate the electronic contribution to the partition function can be calculated explicitly by ignoring quantum fluctuations of $\vec{\psi}$ and, thus, considering it as an external field similarly to how it was done in Ref.[\onlinecite{Wang}]. The thermal operator $\exp(-\beta H)$ can be represented as a product $\prod_i \exp(-\beta H_i)$ over each TLS where $H_i$ is the contribution from the $i$-th TLS.
Then, using the identity $\exp(\vec{B}\vec{\sigma})= \cosh(|\vec{B}|) + \sinh(|\vec{B}) \vec{B}\vec{\sigma}/|\vec{B}|$ for Pauli matrices $\vec{\sigma}$ and real vector $\vec{B}$ and keeping in mind that all the components of $\vec{\sigma}$ are traceless, the contribution to the partition function evaluated with respect to the electronic degrees of freedom becomes
\beq
Z=\exp\left(\beta \mu_0 \int d^2x |\vec{\psi}|^2\right) \prod_i Z_i,\quad Z_i=2\exp(B_i)\cosh (G_i),
\label{Z}
\eeq
where the product is taken over the locations $(\vec{x}_i, z_i)$ of the TLS centers, and the notations
$$
G_i\equiv  \beta \sqrt{(\delta -\mu_0/2)^2 + d^2\phi^2(z_i)|\vec{n}_i\vec{\psi}(\vec{x}_i)|^2},
$$
and
$$B_i\equiv \beta \left[\frac{\mu_0}{2}  - D \phi^2(z_i) |\vec{\psi}(\vec{x}_i)|^2\right]
$$
are introduced.

The orientational relaxation of the TLS is accounted for by the integration $Z_i \to \langle Z_i\rangle_\Omega \equiv \int d\Omega_i Z_i/4\pi$ over the 3D solid angle $\Omega=4\pi$ of the unit vector $\vec{n}_i$. Then, the corresponding contribution to free energy becomes
$\Delta F =-T\sum_i \ln[2e^{B_i}\langle \cosh (G_i)\rangle_\Omega]$. 

In the long wave limit it is reasonable to introduce a coarse grained description according to the rule $\sum_i ... \to \bar{n} \int dz \int d^2 x ...$.  It is also convenient to introduce the dimensionless photonic field according to the substitute  $\vec{\psi}=\psi_0 \vec{\eta}$, where $\psi_0=1/(\beta \phi_0 d)$ and $\phi_0$ is defined in Eq.(\ref{anz}). 
Then, the dimensionless 2D free energy density $f=\Delta F/(T\bar{n}L_z)$ becomes
 \beq
f = \kappa_1 |\vec{\eta}|^2 - \int^{1}_0d\xi \ln \left\langle \cosh (G)\right\rangle_\Omega - \frac{\mu_0}{2T},
\label{f}
\eeq  
where the notations $G=\sqrt{x^2+ g(\xi)|\vec{n}\vec{\eta}|^2}$, $\xi=z/L_z$, $x=\beta (\delta - \mu_0/2)$, $g(\xi)=\sin^2(q_0L_z \xi)$ and
\beq
\kappa_1= \frac{DT}{2d^2} - \frac{T \varepsilon_0}{2d^2 \hbar \omega_0 \bar{n}} \mu_0=\left(\frac{D}{2d^2}-\frac{\varepsilon_0\delta}{d^2\hbar\omega_0\bar{n}}\right)T+\frac{\varepsilon_0x}{d^2\hbar\omega_0\bar{n}}T^2. 
\label{kappa1}
\eeq
are introduced.

The chemical potential $\mu_0$ can be related to the total number of excitations $N_{\rm ex}$ in the system as $N_{\rm ex}= -\frac{\partial \Delta F}{\partial \mu_0}$ giving  the 2D density of the excitations in units of $\bar{n}L_z$ as
\beq
n_{ex}=\frac{\partial f}{2\partial x}=\frac{1}{2}\left[1+ \kappa_2|\vec{\eta}|^2  -x\int^{1}_0 d\xi \frac{\left\langle \frac{\sinh(G)}{G}\right\rangle_\Omega }{\left\langle \cosh(G)\right\rangle_\Omega}\right],
\label{nex}
\eeq
where $\kappa_2= (T^2\varepsilon_0)/(d^2\hbar \omega_0 \bar{n})$ and the density of the normal component of the photonic ensemble is not included (in the limit of speed of light $\to \infty$). [Its contribution will be discussed later in more detail]. 

The total density of the excitations is determined by the balance of energy pumped into the cavity and the rate of the excitations escape.
Introducing the total pumped power $P$ (about $10^2-10^3$mW used in the experiment [\onlinecite{Nature}]), 
the balance condition gives  $N_{\rm ex}\approx\frac{P\tau_e}{\hbar \omega_0}$. 
Thus, Eq.(\ref{nex}) determines the value of the chemical potential $\mu_0$ in terms of $P$.  
Within the mean field approach this equation should be considered together with the condition of the minimum of the free energy $f$. Variation of $f$, Eq.(\ref{f}), with respect to $\vec{\eta}^*$ (in the uniform case) gives
\beq
\kappa_1 \vec{\eta} - \int_0^{1}d\xi   \frac{g(\xi) }{\left\langle \cosh(G)\right\rangle_\Omega} \left\langle \frac{ \sinh( G) (\vec{n}\vec{\eta})\vec{n})}{2G} \right\rangle_\Omega=0 .
\label{eta}
\eeq
Eqs.(\ref{nex},\ref{eta}) represent a full system which allows analyzing the nature of the condensation at time scales shorter than times for typical translational  diffusion of the TLS centers. Below it will be seen that, depending on $n_{ex}$ and the parameters, the condensation can proceed as a continuous or discontinuous transition into photonic condensate characterized by linear polarization. 

At this point it is important to note a significant difference between the system considered here and the exciton-polariton condensate in a crystal  (see in Ref. [\onlinecite{Moskalenko}]). In addition to the absence of the classical orientational degrees of freedom, the induced interaction between photons in the second case is completely different, and it allows for circular polarization as considered in Ref.[\onlinecite{Keeling-08}].

\subsection{Landau expansion}\label{sec:canonical}
The analysis in the limit of the dimensionless photonic field $\vec{\eta} \to 0$ can be conducted within the Landau expansion [\onlinecite{Landau}] of $f$, Eq.(\ref{f}), up to sixth order in $\vec{\eta}$:
\beq
f=f_0+ b_2 |\vec{\eta}|^2 + b_4 |\vec{\eta}|^4 +b_6|\vec{\eta}|^6 +(\tilde{b}_4 + \tilde{b}_6|\vec{\eta}|^2) \vec{\eta}^{\,*\,2} \vec{\eta}^{\,2} ,
\label{LU}
\eeq
where $f_0=-\ln(\cosh(x))$ and the coefficients in terms of the parameter $x$ (defined below Eq.(\ref{f})) are
\beq
b_2&=&\kappa_1 - \frac{\tanh x}{12x}, 
\label{b2} \\
b_4&=&\frac{\tanh^2(x)}{2^63x^2} -\frac{1}{2^55x^2}\left(1-\frac{\tanh(x)}{x}\right),
\label{b4} \\
b_6&=&\frac{5}{2^73 x^3}\left[-\frac{1}{35}\left(\left(1+\frac{3}{x^2}\right)\tanh(x) -\frac{3}{x}\right)-\frac{\tanh^3(x)}{27} +\frac{\tanh(x)}{15}\left(1-\frac{\tanh(x)}{x}\right)\right],
\label{b6}\\
\tilde{b}_4&=&-\frac{1}{2^65x^2}\left(1-\frac{\tanh(x)}{x}\right),
\label{tilb4} \\
\tilde{b}_6&=&\frac{\tanh(x)}{2^89 x^3}\left(1-\frac{\tanh(x)}{x}\right) -\frac{1}{2^87x^3}\left[\left(1+\frac{3}{x^2}\right)\tanh(x) -\frac{3}{x}\right].
\label{tilb6}
\eeq
The forms (\ref{LU}-\ref{tilb6}) should be considered together with Eq.(\ref{nex}).

The global symmetry of $f$ is U(1)$\times$Z$_2$. This can be seen after representing $\vec{\eta}$ as  $\vec{\eta}=\vec{a}+{\rm i} \vec{b}$ in terms of two real 2D vectors $\vec{a}, \vec{b}$. There is a freedom to rotate these vectors in the 2D plane by keeping the same angle between them. This defines the U(1) symmetry.
The product $s_z= \vec{a} \times \vec{b}\neq 0$ determines left or right handed polarization of the condensate. There is the freedom  $s_z \to -s_z$ which represents the Z$_2$ symmetry.    
A minimum of $f$ can be achieved either for $\vec{\eta}=0$ (the normal phase) or for some finite $\vec{\eta}$.

Ignoring the non-linear terms, the mean field condition for the condensation is $b_2=0$ in the expansion (\ref{LU}). This condition does not distinguish between linear and circular polarizations. However, for $b_2<0$ the higher order terms become important. 
 In a uniform case the negative sign of $\tilde{b}_4$ guarantees that $\vec{\eta}$ (in the limit $\vec{\eta} \to 0$) describes linearly polarized photonic condensate -- that is, with $\vec{a},\vec{b}$ being collinear to each other. Thus, it is reasonable to set $\vec{\eta}= \vec{a} \exp(i\varphi)$, where $\varphi$ is some real phase,  in Eq.(\ref{LU}). This gives $f=f_0 +b_2 |\vec{a}|^2 + (b_4+\tilde{b}_4)|\vec{a}|^4 +(b_6+\tilde{b}_6)|\vec{a}|^6$.

\subsection{Phase diagram and the dependence on the pumping rate.}

The parameters controlled experimentally are temperature $T$, dye concentration $\bar{n}$, detuning $\delta$ and $n_{ex}$. The dimensionless form (\ref{LU}) has two parameters only -- $\kappa_1$, Eq.(\ref{kappa1}), and $x$ (defined below Eq.(\ref{f})). It is convenient to consider them as free parameters. The line of IInd order transitions is determined by the condition $b_2=0$ provided $B_4\equiv b_4 + \tilde{b}_4>0$.  As the analysis of Eqs.(\ref{b4},\ref{tilb4}) shows, the latter condition holds for $|x|<x_c, \,\, x_c=1.987(1)$ (given by the condition $B_4=0$). For $|x|>x_c$ the transition becomes of Ist order because $B_4$ changes sign. The transition point is given by $b_2=B^2_4/(4B_6)>0$, provided $B_6=b_6+\tilde{b}_6>0$. Both conditions, $B_4<0,\, B_6>0$ hold in the range $x_c <|x|< 11.25$. 

It is important to note that the Landau expansion describes quantitatively well the discontinuous transition only within the range $x_c< x \leq 2.1$.
At larger $x$ the dimensionless density of photons $\eta^2=|\vec{\eta}|^2$ becomes bigger than unity, and the full expression (\ref{f}) must be used.
 \begin{figure}[!htb]
	\includegraphics[width=0.8 \columnwidth]{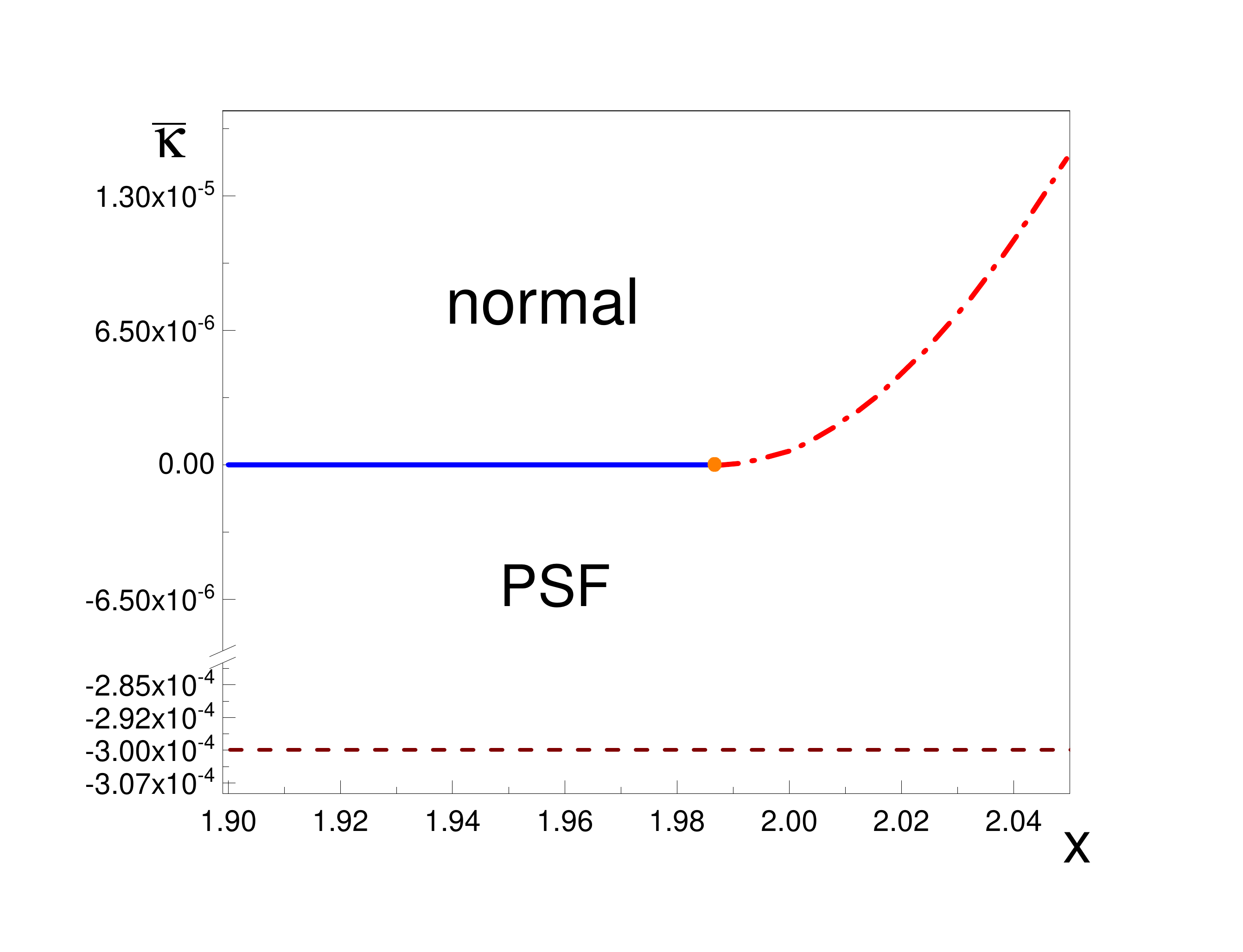}
	\caption{(Color online) The phase diagram in the vicinity of the tricritical point $x=x_c$ separating continuous (solid line) and discontinuous (dot-dashed line) transitions. In the phase labeled as "normal" there is no photonic condensate, that is, $\vec{\eta}=0$ and in the PSF phase  $\vec{\eta}\neq 0$. Here $\bar{\kappa}= \kappa_1 - \tanh(x)/12 x$. The  continuous and discontinuous transition lines are given by the equations $\bar{\kappa}=0, x<x_c$, and $\bar{\kappa}=\frac{B_4^2}{4B_6},\,\, x> x_c$, respectively. The dashed line corresponds to the path along which the data are presented in Fig.\ref{fig3}.}
	\label{fig4}
\end{figure}
\begin{figure}[!htb]
	\includegraphics[width=0.8 \columnwidth]{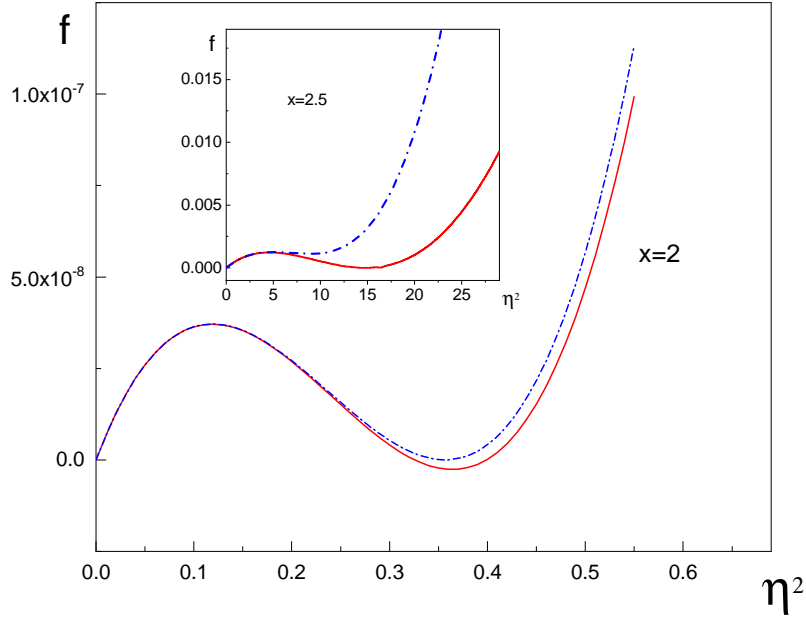}
	\caption{(Color online)  Free energy as a function of the condensate density (shifted to $f=0$ at $\eta=0$) at two points, $x=2$ (main panel) and $x=2.5$ (inset), along the line (dot-dashed) of Ist order transitions, Fig.\ref{fig4}. The continuous and dot-dashed lines represent the full energy (\ref{f}) and the Landau expansion (\ref{LU}), respectively. In the inset the transition point is found for the full energy (\ref{f}).}
	\label{fig2}
\end{figure}
\begin{figure}[!htb]
	\includegraphics[width=0.8 \columnwidth]{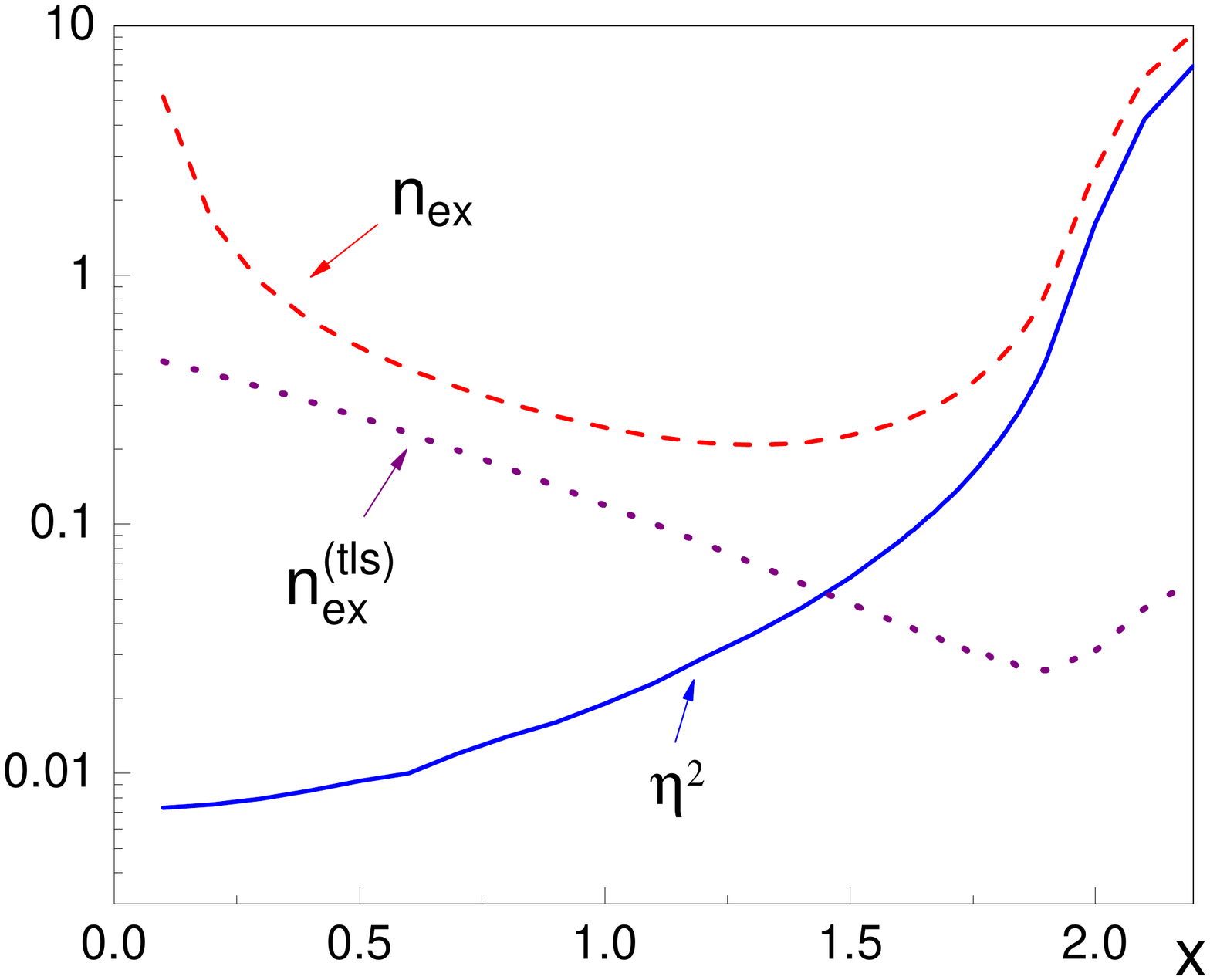}
	\caption{(Color online) Condensate density $\eta^2$, the total excitation density $n_{ex}$ and the density of the excited TLS $n^{(tls)}_{ex}= n_{ex} - \kappa_2 \eta^2$ along the dashed line shown in Fig.\ref{fig4}.}
	\label{fig3}
\end{figure}
The phase diagram 
is shown in Fig.\ref{fig4}. 
Typical dependencies of the free energy on the condensate density $\eta^2$ along the Ist order transition line are shown in Fig.\ref{fig2} for two values of $x>x_c$ -- one which is accurately described within the Landau expansion and the other one for which  the Landau expansion gives only a qualitative description.
For $x$ above the upper limit $x\approx 11.25$, the Landau expansion looses even its qualitative meaning because both $B_4$ and $B_6$ become negative. In reality the transition becomes of strongly Ist order into the values of the condensate density much larger than unity.
This corresponds to large pumping powers, and the analysis of the system must go beyond our model.

Fig.\ref{fig3} represents the condensate (dimensionless) density $\eta^2$, the total density of the excitations $n_{ex}$ and the density of the excited TLS centers in the PSF phase. As mentioned above, $n_{ex}$ carried by excited TLSs can exceed significantly the density of the normal photons close to the condensation threshold. In the case considered here this situation is realized for $x\leq 7$ (see Sec.\ref{sec:normal} for the explanation).
Thus, the condensation condition follows from Eq.(\ref{nex}) as
\beq
n_{ex}=n^{(0)}= \frac{1}{2}\left[1- \tanh(x)\right], 
\label{net}
\eeq
which determines $x$ as a function of $n_{ex}$. [In what follows we will be considering $x>0$ because negative values correspond to the situation of the population inversion $n_{ex}>1/2$]. 

The question is how to translate the phase diagram, Fig.\ref{fig4}, and the dependencies shown in Fig.\ref{fig3} into the experimentally tunable parameters.
Introducing the dimensionless temperature $\tau=T/\delta$ and expressing  it from Eq.(\ref{kappa1}) as 
\beq
\tau=\tau(\kappa_1)=\frac{1}{2x}\left[\gamma_0 +\sqrt{\gamma_0^2+ \gamma_1  x \kappa_1}\right] ,
\label{tau}
\eeq
where the notations $\gamma_0=1- \hbar \omega_0 D\bar{n}/(2\delta \varepsilon_0)$ and $\gamma_1=4\hbar \omega_0 d^2 \bar{n}/( \varepsilon_0\delta^2)$ are introduced, relates temperature to the parameter $\kappa_1$. 
At this point we note that for practical values of the parameters used in the setup [\onlinecite{Nature}] the parameters $\gamma_0$ and $\gamma_1$ obey the conditions $\gamma_0 \approx 1$ and $\gamma_1<<1$. This allows representing $\tau$ as an expansion in $\kappa_1$ as
$\tau\approx \gamma_0/x + \gamma_1 \kappa_1 /(4\gamma_0)$ up to the first order term in $\kappa_1$ only from Eq.(\ref{tau}). Thus, the other physical parameter controlling the phases of light is the deviation 
of temperature $\bar{\tau}=  \gamma_1 \kappa_1 /(4\gamma_0)$ from $\tau=\gamma_0/x \approx 1/x$, and the vertical axis in Fig.\ref{fig4} is essentially given by $\bar{\tau} <<1/x$.

Eq.(\ref{net}) is constrained by the requirement that the free energy has only the trivial solution $\eta=0$ at its minimum. Within the validity of the Landau expansion this corresponds to $b_2\geq 0$ for $0<x<x_c$ in the range of the continuous transition and $b_2 \geq  b^{(c)}_2=B_4^2/(4B_6)$ slightly above the tricritical point $x_c$ -- in the domain of the discontinuous transition.

Eq.(\ref{b2}) gives the condition for the continuous transition as $\kappa_1 = \kappa_c=\tanh(x)/12 x$. According to the definition (\ref{tau}) the temperature along this line becomes  
\beq
\tau = \tau_{\rm II}= \tau(\kappa_1)\approx \frac{\gamma_0}{x} + \frac{\gamma_1}{48 \gamma_0} \frac{\tanh(x)}{x}, \,\, \, 0<x<x_c.
\label{II}
\eeq 
For $x>x_c$ the boundary for the discontinuous transition within the Landau expansion is given by $\kappa_1=\frac{\tanh(x)}{12x} + \frac{B_4^2}{4B_6}$. This defines the temperature along this line as
\beq
\tau = \tau_{\rm I}=\tau(\kappa_1)\approx \frac{\gamma_0}{x} +\frac{\gamma_1}{48 \gamma_0}\left(\frac{\tanh(x)}{x} + \frac{3 B_4^2}{B_6}\right) ,\,\,\, x>x_c.
\label{I}
\eeq 
In both equations (\ref{II},\ref{I}) the value of $x$ is defined by $n_{ex}$ (that is, by the pumping power) from Eq.(\ref{net}).

As mentioned already, while the continuous line (\ref{II}) is well described within the Landau expansion, the solution (\ref{I}) is close to the actual curve only for $x\leq 2.1$. Above this value, the order parameter $\eta$ becomes significantly larger than unity along the transition line, and the full expression for the free energy (\ref{f}) must be used to obtain the transition line.


 \subsection{The role of normal photons and comparison with the experiment}\label{sec:normal}

At this point, let's discuss the accuracy of the approximation where the normal component of photons $n_n$ can be ignored.
At the (ideal) BEC transition this density can be estimated in 2D as
\beq
n_{n} = 2\int \frac{d^2k}{(2\pi)^2} \frac{1}{\exp(\hbar^2k^2/(2mT))-1} \approx \frac{T\omega_0}{\pi \hbar c^2} \ln(L/L_z), 
\label{Nn}
\eeq 
where the log-divergence is cut off by the system size. For the typical sizes and temperatures of the experiment [\onlinecite{Nature}] we find $n_n \sim (10^{11}-10^{12}) {\rm m}^{-2}$. 
These values should be compared with the density of excitations carried by the dye molecules in the normal phase. 
For the power $P$  (about $10^2-10^3$mW used in the experiment [\onlinecite{Nature}]) the total number of excitations $N_{ex} \sim 10^8-10^{10}$.
In the active volume about  $2\mu$m$ \times 100\mu$m$ \times 100\mu$m [\onlinecite{Nature}] this implies the 2D density of the excitations $ (10^{16}-10^{18})$m$^{-2}$.
These numbers should be compared with the 2D density  $\bar{n}L_z$ of the TLS centers. In  the system [\onlinecite{Nature}] the fractional concentration of the dye molecules is about $10^{-3}$. Thus,  $\bar{n} L_z \sim 10^{18}$m$^{-2}$.  
The normal photons can be safely ignored if $n_n << n_{ex}\bar{n}L_z$ for a given value of $x$ as determined by Eq.(\ref{net}). 
The above estimates  show that  this condition is violated for $x \geq 7$ which is well beyond the vicinity of the tricritical point. Thus, the approximation ignoring $n_n$ in our analysis is well justified.

It is useful to estimate where the actual experiment [\onlinecite{Nature}] is on the phase diagram. For the parameters  relevant to the experiment, the quantity $x$ is of the central importance. For the relevant values of $\gamma_0, \gamma_1$ it is given by the ratio $x\approx \delta /T $ of the half-detuning $\delta$ to the temperature $T$.
The detuning is related to zero phonon line energy. Using the approach [\onlinecite{Hughes}] its value can be estimated as $\delta \approx \hbar \omega_0 \delta \lambda/\lambda $ with $\delta \lambda \approx 20$nm, where $\lambda$ stands for the wavelength corresponding to the frequency $\omega_0$ and $\delta \lambda$ is the difference with the corresponding wavelength of the zero phonon line. This gives
$x\approx 3.0$ for $T\approx 300$K which is in the domain of strong Ist order transition. In this region the dimensionless photon density $\eta^2 \approx 20$ according to Landau expansion and $\eta^2 \approx 30$ as follows from the full energy (\ref{f}). 

It is important to note that in the experiment [\onlinecite{Nature}] the condensate density does not show any features typical for Ist order transition. The question is why so. Our analysis has been conducted in the thermodynamical limit of the uniform system. In the experiment the condensate has finite extension about 14$\mu$m due to the trapping potential. Thus, our analysis can be applied directly to the experiment [\onlinecite{Nature}] only if the local density approximation holds.   This means that a typical scale $R_n$ for the nucleation occurring during Ist order transition must be much smaller than the condensate size. This scale is determined by the extension of the domain wall separating normal and PSF phases. In order to obtain the corresponding solution the part (\ref{U0}) must be included into the analysis. Then, we find $R_n \approx 10-15\mu$m for $x=3$ which is comparable to the condensate size. This implies that the local density approximation cannot be applied to the experiment [\onlinecite{Nature}]. That is, the discontinuous nature of the transition cannot be determined in so small samples.

\section{Grand canonical description: phase separation effect}\label{GC}

At longer times, translational rearrangement of the TLS positions should take place due to their collective interaction with the photonic field. This effect can be described within the grand canonical approach with respect to the number of TLS centers $N_{TLS}$ by introducing another chemical potential $\mu_1$ -- now for the total TLS number, with the term   $ - \mu_1 N_{TLS}$ added to the total Hamiltonian.
Then, the product over the TLS positions in Eq.(\ref{Z}) is to be replaced by the product over  the spatial cells covering the resonator volume at density $n_0 \approx a^{-3}$ and  each containing either none or one TLS molecule of the linear size $a \sim 1$ nm. Then, the partition function averaged over the TLS degrees of freedom becomes 
\beq
Z^{(gc)}=\prod_i Z^{(gc)}_i,\,\,\, Z^{(gc)}_i= 1 + 2e^{\beta \mu_1 + B_i} \left\langle \cosh( G_i))\right\rangle_\Omega,
\label{Zfull}
\eeq
where now the product $\prod_i ...$ is taken over all the spatial cells covering the whole volume, and the notations $B_i, G_i$ have been introduced below Eq.(\ref{Z}). It is instructive to compare $Z^{(gc)}_i$ with $Z_i$ from Eq.(\ref{Z}). The first and the second terms in $Z^{(gc)}_i$ from Eq.(\ref{Zfull}) correspond, respectively, to zero and one TLS present in the i-th cell. The corresponding contribution to the uniform part of the free energy (grand canonical potential) $\Delta F_{gc}=-T\sum_i \ln Z^{(gc)}_i$ becomes
 \beq
f_{gc} = \kappa_{1gc} |\vec{\eta}|^2 - \int^{1}_0d\xi \ln\left(1+Ye^{-\gamma_D g(\xi) |\vec{\eta}|^2} \left\langle \cosh(G)\right\rangle_\Omega\right),
\label{fgc}
\eeq  
where $f_{gc} = \Delta F_{gc}/(Tn_0 L_z)$,  $\gamma_D=T D/d^2$, and the coarse grained description following the rule $\sum_i ... \to n_0\int dz \int d^2 x ...$ is introduced; $\kappa_{1gc}=-\varepsilon_0\mu_0T/(2\hbar \omega_0 d^2 n_0)$; $Y=2\exp(\beta(\mu_1 + \mu_0/2))$;  $G$ and $\vec{\eta}$,$x$ variables were introduced in Sec.\ref{canonical} above and below Eq.(\ref{f}).  

The value of $N_{TLS}$ follows from $N_{TLS}= - \partial \Delta F_{gc}/\partial \mu_1$. It gives the 2D TLS density in units of $n_0L_z$ as
\beq
 \bar{n} =  \int_0^1 d\xi \frac{Ye^{-\gamma |\vec{\eta}|^2}\langle \cosh(G)\rangle_\Omega}{1+Ye^{-\gamma |\vec{\eta}|^2}\langle \cosh(G)\rangle_\Omega}.
\label{ntls}
\eeq
Without the PSF this relations gives $\bar{n}=Y\cosh(x)/(1+Y\cosh(x))$. In the case $\bar{n} <<1$, it is reasonable to consider the limit $Y<<1$ which justifies  keeping only the linear in $Y$ term in $f_{gc}$, Eq.(\ref{fgc}) ( $x$ values are limited by $x\sim 1-2$), as
\beq
f_{gc} =  \kappa_{1gc} |\vec{\eta}|^2 - Y\int^{1}_0d\xi  e^{-\gamma |\vec{\eta}|^2} \left\langle \cosh(G)\right\rangle_\Omega.
\label{fgcY}
\eeq  
Accordingly, $\bar{n}$ from Eq.(\ref{ntls}) simplifies as 
\beq
\bar{n}=Y \int_0^1 d\xi  e^{-\gamma |\vec{\eta}|^2} \langle \cosh(G)\rangle_\Omega.
\label{nbar}
\eeq
Comparing the above two relations it is useful to note that in the grand canonical ensemble the dimensionless free energy and the TLS concentration obey a very simple relation $f_{gc} = \kappa_{1gc} |\vec{\eta}|^2 - \bar{n}$. Thus, if there is a discontinuous transition with respect to the order parameter $\vec{\eta}$, there
is the bimodality in the TLS density as well. This automatically implies the phase separation effect -- TLS density as well as photonic condensate both become spatially non-uniform. This features a coexistence of the normal and the PSF phases.

Within the same accuracy the 2D density of the excitations follows from $N_{ex}=- \partial \Delta F_{gc}/\partial \mu_0$ (in units of $n_0L_z$) as
\beq
n_{ex}=\kappa_{2gc}|\vec{\eta}|^2 +\frac{Y}{2}\int^{1}_0 d\xi  e^{-\gamma |\vec{\eta}|^2}\left\langle\cosh(G) -   \frac{x\sinh(G)}{G}\right\rangle_\Omega, 
\label{nexc}
\eeq
where $\kappa_{2gc}=\varepsilon_0 T^2/(2 \hbar \omega_0 d^2 n_0)$.
It is worth mentioning that in the normal phase, $\vec{\eta}=0$, this relation transforms into the form (\ref{net}) if $n_{ex}$ is expressed in units 
of the actual TLS 2D density $\bar{n}L_z=n_0 Y\cosh(x) L_z$.

\subsection{Landau expansion in the grand canonical description.}
The tricritical point (separating the continuous and discontinuous transitions) in the phase diagram exists in the grand canonical description as well. It can also  be found by using the Landau expansion of $f_{gc}$, Eq.(\ref{fgcY}), in $\vec{\eta}$ up to the 6th order. The difference with the analysis conducted in  Sec.\ref{canonical} is in the consequence of the discontinuous formation of the PSF. In the first case, no phase separation takes place -- only the adjustment of the molecular orientations proceeding  together with the condensate formation. Here the transaltional motion of the dye molecules is included into the consideration too, and this results in the phase separation -- occurring simultaneously with the Ist order transition.

The Landau expansion of the grand canonical free energy,  Eq.(\ref{fgcY}), becomes (up to a constant)
\beq
f_{gc}=  b_{2gc} |\vec{\eta}|^2 + b_{4gc} |\vec{\eta}|^4 +b_{6gc}|\vec{\eta}|^6 +(\tilde{b}_{4gc} + \tilde{b}_{6gc}|\vec{\eta}|^2) \vec{\eta}^{\,*\,2} \vec{\eta}^{\,2},
\label{LGC}
\eeq
where the coefficients are
\beq
b_{2gc}&=&\kappa_{1gc} +\left(\frac{\gamma_D}{2} - \frac{\tanh x}{12x}\right) \cosh(x) Y, 
\label{b2gc} \\
b_{4gc}&=&\frac{3}{48}\left[ - 3\gamma^2_D  + \gamma_D \frac{\tanh(x)}{x} - \frac{1}{10x^2}\left(1- \frac{\tanh(x)}{x}\right) \right]\cosh(x) Y,
\label{b4gc} \\
b_{6gc}&=&\frac{5}{192}\Big[2 \gamma^3_D  - \gamma_D^2\frac{\tanh(x)}{x} + \frac{\gamma_D}{5x^2} \left(1-\frac{\tanh(x)}{x}\right) - 
\nonumber \\
 &&\frac{1}{70x^3}\left(\left(1+\frac{3}{x^2}\right)\tanh(x) -\frac{3}{x}\right)\Big]\cosh(x) Y,
\label{b6gc}\\
\tilde{b}_{4gc}&=&-\frac{1}{320x^2}\left(1-\frac{\tanh(x)}{x}\right)\cosh(x) Y,
\label{tilb4gc} \\
\tilde{b}_{6gc}&=&\frac{1}{384x^2}\left[\gamma_D \left(1-\frac{\tanh(x)}{x}\right) -\frac{3}{14x}\left(\left(1+\frac{3}{x^2}\right)\tanh(x) -\frac{3}{x}\right)\right]\cosh(x) Y,
\label{tilb6gc}
\eeq
and the parameter $\gamma_D$ has been introduced below Eq.(\ref{fgc}); $x$ has the same meaning as in the expansion in Sec.\ref{sec:canonical}.

Similar to the situation considered in Sec.\ref{canonical}, the condensate is characterized by linear polarization because $\tilde{b}_{4gc}<0$.
Then, using the relation $\vec{\eta}^{*\,2}\vec{\eta}^2=|\vec{\eta}|^4$ with $\vec{\eta}=\vec{a}\exp(i\varphi)$, where $\vec{a}$ is a real vector, in $f_{gc}$, the condition for the tricritical point becomes $B_{4gc}\equiv b_{4gc} + \tilde{b}_{4gc} =0$ provided $B_{6gc}\equiv b_{6gc} + \tilde{b}_{6gc} >0$. 
In contrast to the case considered in Sec.\ref{canonical} with only one tricritical point (at $x=x_c$), here there is a line of such points
existing for $0<x<x_c$. It can be expressed as the relation 
\beq
\gamma_D=\gamma_\pm (x)=\frac{\tanh(x)}{6x}\left[1\pm \sqrt{1-\frac{9(1-\tanh(x)/x)}{5\tanh^2(x)}}\right],\,\,\, 0<x<x_c,
\label{tri}
\eeq
between the parameters $\gamma_D,x$, with the sign $\pm$ correlated. 
It is worth mentioning that the equation for $x_c$ above which the real solutions (\ref{tri}) vanishes coincides with the solution $B_4=0$ from Eqs.(\ref{b4},\ref{tilb4}).

\begin{figure}[!htb]
	\includegraphics[width=0.8 \columnwidth]{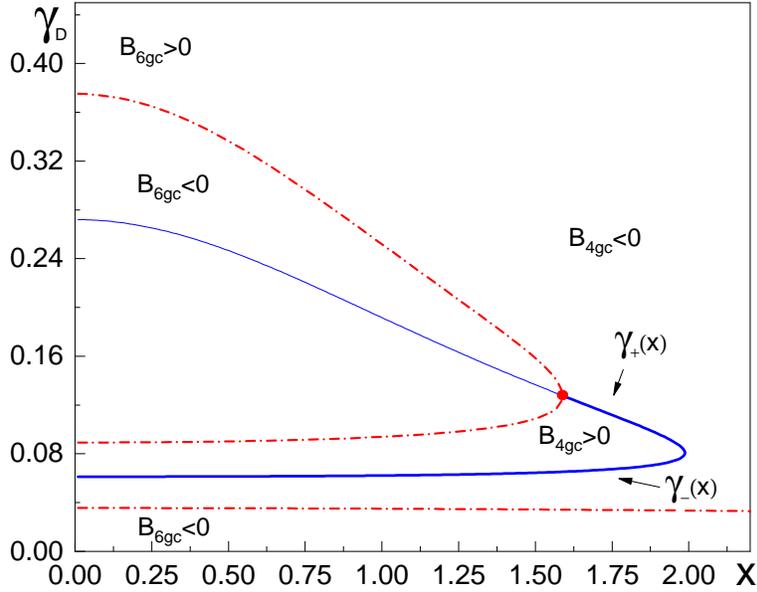}
	\caption{(Color online) The domains of the transitions in the grand canonical ensemble for photons and dye molecules. The solid line shows tricritical points where $B_{4gc}=0$. Its thick portion accounts for the set of stable tricritical points separating IInd order (where $B_{4gc}>0, B_{6gc}>0$) and Ist order ($B_{4gc}<0, \, B_{6gc}>0$) transitions. The dot-dashed lines correspond to the roots of the equation $B_{6gc}=0$. These lines break the whole field into three regions. Inside the regions where $B_{6gc}<0$ the Landau expansion becomes invalid, and these regions are beyond the scope of our model. The applicability of the Landau expansion is practically limited to the vicinity of the solid thick line.} 
	\label{fig5}
\end{figure}

The condition for the continuous transition $ b_{2gc}=0,\,\, B_{4gc}>0$ is satisfied
if $\gamma_-(x) <\gamma_D< \gamma_+(x)$, and the first order transition happens outside this domain. The stability condition for the tricritical point
 $B_{6gc}>0$ is satisfied  along the line $\gamma_D=\gamma_-(x)$ for $ 0<x<x_c$ and for  $1.59(2) <x<x_c$ along $\gamma_D=\gamma_+(x)$. The full diagram of the domains of the transitions is shown in Fig.\ref{fig5} in terms of the variables $\gamma_D \sim T$ and $x$ .
\begin{figure}[!htb]
	\includegraphics[width=0.8 \columnwidth]{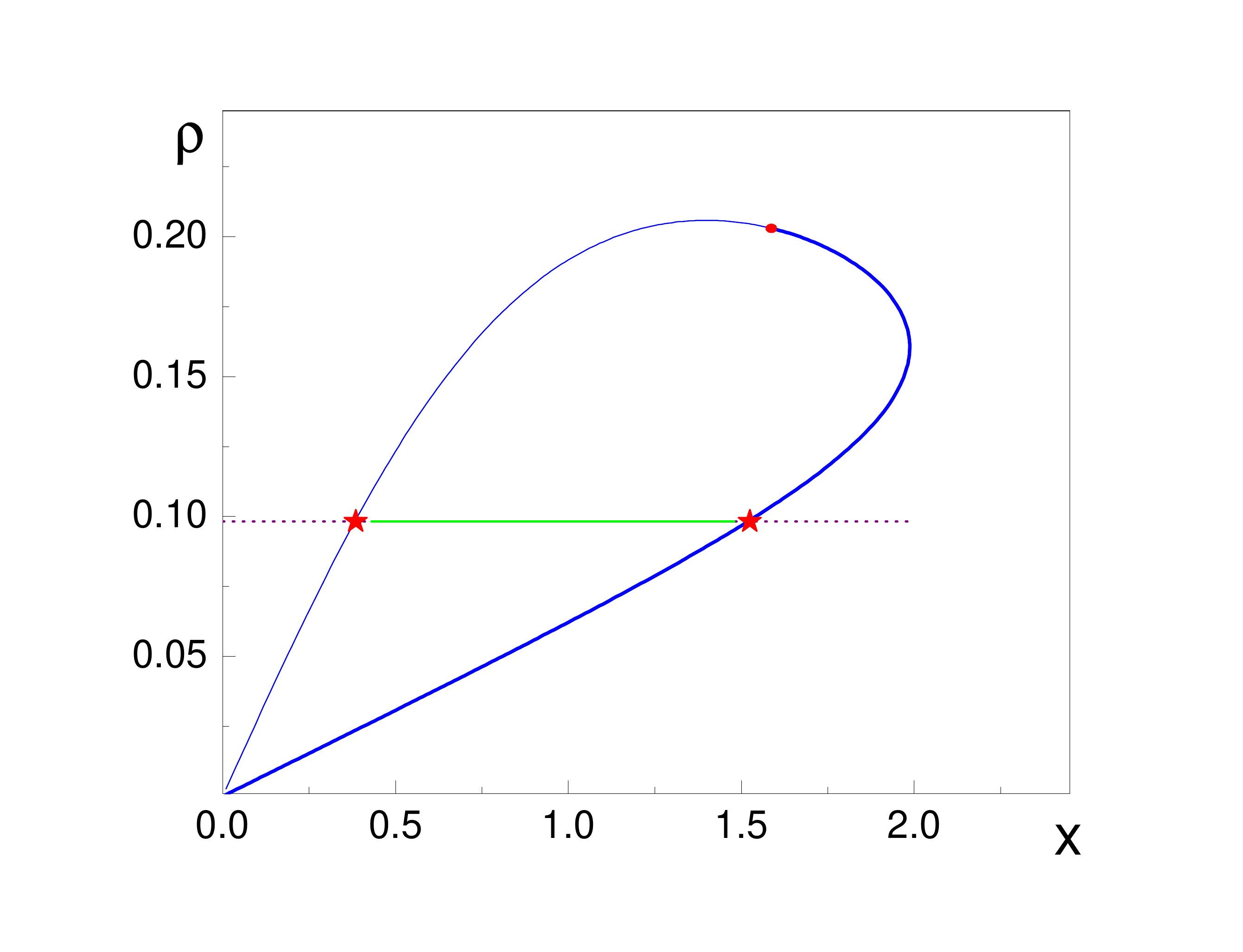}
	\caption{(Color online) The range of the parameter $\rho$ where IInd order transition can occur -- inside the area surrounded by the solid curved line. As an example, the horizontal solid straight line $\rho\approx 0.1$ shows the corresponding values of $x$. The symbol (star) on the right end of the line labels the corresponding (stable) tricritical point ($x_t\approx 1.525$). For $x>x_t$ (along the dotted line) the transition becomes of Ist order. The symbol (star) on the left end indicates the ending of the validity of the Landau expansion. The meanings of the solid thick and thin lines are the same as in Fig.\ref{fig5}. } 
	\label{fig6}
\end{figure}

It is important to note that the condition for the IInd order transition $b_{2gc}=0$ coincides with the one, $b_2=0$, from Sec.\ref{canonical}. Thus, the condition in terms of the reduced temperature $\tau$ is given by Eq.(\ref{II}), with $x$ determined from Eq.(\ref{net}). The requirement that $\gamma_D$ belongs inside the region surrounded by the solid line in Fig.\ref{fig5}, or
$\gamma_-(x)< \gamma_D < \gamma_+(x)$, where $\gamma_\pm$ is given in Eq.(\ref{tri}) can be represented as 
\beq
x\gamma_-(x) < \frac{D\delta}{d^2} <x \gamma_+(x),
\label{IIgc}
\eeq 
where the definition $\gamma_D= (D\delta/d^2)\tau$  and $\gamma_0 \approx 1$ and $\gamma_1<<1$ are used, and, thus, the reduced temperature $\tau$ is taken as $\tau\approx 1/x$.
This implies that the continuous transition can occur if the parameter $\rho=D\delta/d^2$ stays in the range $0<\rho <0.203$ (see in Fig.\ref{fig6}). The tricritical point $x=x_t$ is determined from the equation $x\gamma_-(x)=\rho$, and $x\gamma_+(x)=\rho$ defines the ending point beyond which the applicability of the Landau analysis ends. These two solutions are shown in Fig.\ref{fig6} by stars for one choice of $\rho$.

It is worth noting that the phase separation does not require that the free energy (\ref{LGC}) becomes zero for $\eta \neq 0$. It is enough 
that $f_{gc}$ acquires a metastable solution (spinodal). This happens at $b_{2gc} = B^2_{4gc}/(3B_{6gc})$ (as compared to $b_2=B^2_4/(4B_6)$ in the case considered in Sec.\ref{canonical}).  The actual fractions of the coexisting phases are to be determined by the Maxwell construction.

\section{Discussion}\label{Disc}
The main result of this paper is that, depending on the parameters, the photonic Bose-Einstein condensation can proceed as continuous or discontinuous transition. This happens because of the involvement of the mechanical degrees of freedom  of the dye molecules -- their rotations and translations. At the time scale up to few ns only the orientations of the molecules control the nature of the transition. At longer times, the translational dynamics becomes important too, and this leads to the phase separation effect which features coexistence of the PSF and normal phases.
There are tricritical points on the phase diagram in terms of the physically tunable parameters. Detecting such a point represents an important task for the experiment. One of the typical features indicative of the discontinuity is hysteresis. It should be observable by cycling the pump power in the corresponding region of the phase diagram. 

As the analysis of Sec.\ref{sec:normal} shows, it is impossible to detect the discontinuity in the settings of the experiment [\onlinecite{Nature}] unless the system is taken farther away from the tricritical point $x\approx 2$. Practically, this can be achieved by increasing either the size of the condensate or the value of the detuning $\delta$. For example, by increasing $\delta$ by a factor of 2, the minimal scale for the nucleation decreases by a factor of 2. This may result in observing the hysteresis.  

There is also a question about the nature of the continuous transition (for $x <x_c \approx 2$). According to the mean field analysis it should feature U(1) universality of 2D vector describing linearly polarized condensate. However, there is a degeneracy at the transition with respect to the circularly polarized condensate. Thus, it is not excluded that the actual universality is characterized by higher symmetry or fluctuation induced Ist order transition. The analysis of such options requires going beyond the mean field approximation.  At this point it is important to notice that experimentally the PSF and the normal cloud polarization was found to be determined mostly by the polarization of the pump and hidden structural anisotropy of the resonator [\onlinecite{Greveling}]. This situation corresponds to the presence of some effective external field which smears out the actual transition. Thus, the analysis of the universality must include this option.  To tune the system into this region, the value of the detuning should be decreased by at least a factor of 1.5-2. To perform proper scaling in the experiment there should be the option to create samples of variable size.  

At this point it is useful to note the differences between the present analysis of the condensate polarization and the one conducted in Ref.[\onlinecite{Moodie-2017}]. 
The focus of Ref. [\onlinecite{Moodie-2017}]  is on the early stages of the kinetics of the condensate formation -- before the quasi-condensate appears.  The present analysis focuses on the equilibrium phases of photons and dye molecules where the off diagonal order is essential.

In our analysis the interaction between TLS centers mediated by the near zone photons was not taken into account. Such an interaction, as considered in Refs.[\onlinecite{Sela-14,SFY16}], introduces another option -- a strongly correlated state of TLSs without any photonic condensation. The question to answer is how these two states -- photonic condensate and the TLS phase -- affect each other.   

In our analysis the only mechanism for inducing the effective interaction between photons in the condensate is due to stimulated virtual electronic transitions of the dye molecules [\onlinecite{Wang}]. There is also another mechanism related to the non-radiative transitions from the excited molecular levels leading to heating of the condensate and inducing the so called thermal lensing (see in Ref.[\onlinecite{Nature}]). Strength of this effect depends on thermal conductivity of the system and the geometry. Furthermore, this effect creates non-uniformity following the profile of the thermal gradient.  Full analysis of the actual experiment should include both mechanisms.  

 {\it Acknowledgments}.
 This work was supported by the National Science Foundation under the grant DMR-1720251.  We thank   Center for Theoretical Physics of Complex Systems, Institute for Basic Science Korea, for hospitality. A.B.K. also acknowledges useful discussion of dye lasers with A. Gorokhovsky.




\begin{thebibliography}{99}
\bibitem{note} The actual cavity in Ref.[\onlinecite{Nature}] has a curvature in order to induce photonic trap and to reduce the low dimensional fluctuations. Here we consider the thermodynamic limit in the uniform 2D geometry within the mean field approach which ignores such fluctuations.
\end{thebibliography}
\end{document}